\documentclass[final,5p,twocolumn,number,times,10pt]{elsarticle}
\biboptions{sort&compress,comma}
\usepackage[T1]{fontenc}

\makeatletter
\usepackage{pst-node}\usepackage{lscape}\usepackage{amsfonts}\usepackage{amsbsy}\usepackage{graphics}\usepackage{lscape}\usepackage{mathrsfs}\usepackage{dsfont}\@ifundefined{definecolor}
 {\usepackage{color}}{}

\newcommand{\rM}{\mathrm{M}}

\newcommand{\Tr}{\mathrm{Tr}}

\newcommand{\bea}{\begin{eqnarray}}
\newcommand{\eea}{\end{eqnarray}}

\newcommand{\hsp}{\hspace{-.25cm}}
\newcommand{\hspp}{\hspace{-.075cm}}
\newcommand{\nn}{\nonumber \\}
\newcommand{\nnn}{\nonumber }

\def\slash#1{\setbox0=\hbox{$#1$}  
   \dimen0=\wd0     
   \setbox1=\hbox{/} \dimen1=\wd1  
   \ifdim\dimen0>\dimen1   
      \rlap{\hbox to \dimen0{\hfil/\hfil}} 
      #1     
   \else     
      \rlap{\hbox to \dimen1{\hfil$#1$\hfil}} 
      /      
   \fi}      %

\@ifundefined{showcaptionsetup}{}{%
 \PassOptionsToPackage{caption=false}{subfig}}
\usepackage{subfig}
\makeatother

\makeatother

\begin{document}

\begin{frontmatter}
\title{Final state interactions and the transverse structure of the pion
using non-perturbative eikonal methods}

\author[label1,label2]
{Leonard Gamberg\corref{cor1}}
\cortext[cor1]{corresponding author}
\ead{lpg10@psu.edu}
\author[label3]{Marc Schlegel}
\ead{schlegel@jlab.org}
\address[label1]{Division of Science, Penn State Berks, Reading, PA 19610, USA}
\address[label2]{Institute for Nuclear Theory, Seattle, WA 98103, USA}
\address[label3]{Theory Center, Jefferson Lab, Newport News, VA 23606, USA
}

\begin{abstract}
In the factorized picture of semi-inclusive hadronic processes
the naive  time reversal-odd parton distributions exist by virtue
of the gauge link which renders it color gauge invariant.
 The link characterizes the dynamical effect of
initial/final-state interactions of the active parton due  soft
gluon exchanges with the target remnant. Though these  interactions
are non-perturbative,   studies of
 final-state interaction have been approximated by
 perturbative one-gluon approximation in Abelian models.
We include higher-order gluonic contributions from the gauge link 
by applying 
non-perturbative eikonal methods 
incorporating color degrees of freedom 
in a calculation of  the Boer-Mulders
function of the pion.
Using this framework  we explore 
under what conditions the Boer Mulders function  can be described in
terms of factorization of final state interactions and a spatial distribution
in impact parameter space.
\end{abstract}


\end{frontmatter}

\section{Introduction}
\label{intro}
Over the past two decades the transverse partonic structure of hadrons
has been the subject of a great deal of theoretical and experimental
investigation. Central to these studies are the early observations
of large transverse single spin asymmetries (TSSAs) in 
inclusive hadron production from proton proton scattering over a wide
range of beam energies~\cite{Bunce:1976yb,Dragoset:1978gg,Antille:1980th,Adams:1991cs}. Recently TSSAs have been observed in lepton-hadron semi-inclusive
deep inelastic scattering (SIDIS) at the COMPASS~\cite{Schill:2008ra}
and HERMES~\cite{Airapetian:1999tv,Airapetian:2004tw} experiments
and at Jefferson Lab~\cite{Avakian:2003pk,Avakian:2005ps}, as well
as in inclusive production of pseudo-scalar mesons in proton-proton
collins at RHIC~\cite{Adams:2003fx,Adler:2005in,Arsene:2008mi,Abelev:2008qb}.
While the naive parton model predicts
that transverse polarization effects are 
trivial in the helicity limit~\cite{Kane:1978nd}, 
it has been 
demonstrated~\cite{Efremov:1984ip,Efremov:1981sh,Qiu:1991pp,Qiu:1991wg} 
that soft gluonic and fermionic pole contributions to
multiparton correlation functions result in non-trivial twist-three
transverse polarization effects.   In addition
theoretical work on transversity ~\cite{Ralston:1979ys,Jaffe:1991kp,Collins:1992kk}
indicated that transverse polarization effects can appear at leading
twist. Two explanations to account for TSSAs in QCD have emerged which
are based on the twist-three~\cite{Qiu:1991pp,Qiu:1991wg} and 
twist-two~\cite{Sivers:1989cc,Collins:1992kk,Anselmino:1994tv,Mulders:1995dh,Boer:1997nt}
approaches. Recently, a coherent picture has emerged which describes
TSSAs in a kinematic regime where the 
two approaches are expected to have a common description~\cite{Boer:2003cm,Ji:2006ub,Ji:2006br,Bacchetta:2008xw}.

In the factorized picture of SIDIS~\cite{Mulders:1995dh,Ji:2004wu}
at small transverse momenta of the produced hadron
$P_T\sim k_T << \sqrt{Q^2}$
the Sivers effect describes a twist-two  
transverse target
spin-$S_{T}$ asymmetry  through the {}``naive'' time reversal odd (T-odd) 
structure, $\Delta f(x,\vec{k}_{T})\sim S_{T}\cdot(P\times\vec{k}_{T})f_{1T}^{\perp}(x,k_{T}^{2})$~\cite{Sivers:1989cc,Sivers:1990fh} where  $\sqrt{Q^2}$ is the hard scale, $k_{T}$
 is the quark intrinsic transverse momentum, and $P$ is
the momentum of the target. 
For an unpolarized target with transversely polarized quarks-$s_{T}$,
the Boer-Mulders function~\cite{Boer:1997nt} is given by 
$\Delta h(x,\vec{k}_{T})\sim s_{T}\cdot(P\times\vec{k}_{T})h_{1}^{\perp}(x,k_{T}^{2})$.  Dynamically, T-odd-PDFs emerge from the gauge link structure
of the multi-parton quark and/or gluon correlation functions~\cite{Brodsky:2002cx,Collins:2002kn,Belitsky:2002sm,Boer:2003cm}
which describe initial/final-state interactions (ISI/FSI) of the active
parton via soft gluon exchanges with the target remnant. 

Many studies have been performed to model the T-odd PDFs in 
terms of the FSIs  where soft gluon rescattering 
is approximated by perturbative 
one-gluon exchange in Abelian models~\cite{Brodsky:2002cx,Ji:2002aa,Goldstein:2002vv,Boer:2002ju,Gamberg:2003ey,Gamberg:2003eg,Bacchetta:2003rz,Lu:2004hu,Gamberg:2007wm,Bacchetta:2008af}.
We go beyond this approximation by 
applying  non-perturbative eikonal methods to 
calculate higher-order gluonic contributions 
from the gauge link while also taking into 
account color.

In the context of these 
higher order contributions 
we  perform a quantitative study of approximate relations
between TMDs and GPDs. In particular, we 
explore under what conditions the 
T-odd PDFs can be described via factorization of FSI and spatial
distortion of impact parameter space PDFs~\cite{Burkardt:2002hr}.
While such relations
are fulfilled from lowest order contributions in  field-theoretical
spectator models \cite{Burkardt:2003je,Meissner:2007rx} a model-independent
analysis of 
generalized parton correlation functions (GPCFs)~\cite{Belitsky:2003nz}
indicates that the Sivers function and the helicity flip GPD $E$
are projected from independent GPCFs. A similar result holds for the
Boer-Mulders function for a spin zero target \cite{Meissner:2008ay}.
However for phenomenology, it is essentially unknown  whether the
proposed factorization  is a good approximation. 
Here we focus on the transverse structure
of the pion in terms of the impact parameter GPD  $H_1^\pi$,  
and  the Boer Mulders function for which there are very few studies.
Recent lattice calculations~\cite{Brommel:2007xd}  
indicate that the spatial asymmetry of
transversely polarized quarks in the pion is quite similar in magnitude
to that of quarks in the nucleon which lends
supports the findings in~\cite{Burkardt:2007xm}.
Further understanding of the Boer-Mulders function for the pion may 
provide insight into the explanation of large $\cos2\phi\sim h_{1}^{\perp\,\pi}\otimes h_{1}^{\perp}$
azimuthal asymmetry (AA) observed in unpolarized $\pi^- p$ Drell-Yan 
scattering
~\cite{Falciano:1986wk,Guanziroli:1987rp,Conway:1989fs}. This work  
also has direct impact on studies of AAs and TSSAs
in unpolarized and polarized  $\pi N$ Drell-Yan experiments
proposed by the COMPASS collaboration. In the latter case the TSSA
is sensitive to the the nucleon's transversity through the convolution
of $h_{1}^{\perp\,\pi}\otimes h_{1}$.

\section{T-odd PDFs, Gluonic Poles and The Lensing Function}

The field-theoretical definition of transverse-momentum dependent
(TMD) parton distributions in terms of hadronic matrix elements of
quark operators serves as the starting point of our analysis. 
A classification
of TMDs for a spin-1/2 hadron with momentum $P$ and spin $S$ was
presented in Refs.~\cite{Mulders:1995dh,Goeke:2005hb,Bacchetta:2008xw}. 
In an analogous manner, it is straight-forward to obtain the TMDs for a 
spin-0 particle from the correlator for a pseudo-scalar target. 
One encounters two leading twist TMDs for a pion, the distribution for unpolarized quarks $f_1$
and the distribution of transversely polarized quarks $h_1^{\perp\pi}$, 
the Boer-Mulders function.
Adopting the infinite-momentum frame where the hadron moves relativistically
along the positive $z$-axis such that the target momentum $P$ has
a large plus component $P^{+}$ and no transverse component we use
the light cone components of a 4-vector $a^{\pm}=1/\sqrt{2}(a^{0}\pm a^{3})$,
$a^\mu=(a^-,a^+,a^\perp)$. The Boer-Mulders function, defined for SIDIS reads 
\begin{eqnarray}
2\epsilon_T^{ij}k_T^j h_1^{\perp}(x,\vec{k}_{T}^2) \hsp&=\hsp& m_\pi \int\frac{dz^{-}d^{2}z_{T}}{2(2\pi)^{3}}\,\,\mathrm{e}^{ixP^{+}z^{-}-i\vec{k}_{T}\cdot\vec{z}}
\nn
&&\hspace{-2.0cm}
\times\langle P|\,\bar{q}_{j}(0)\,[0\,;\,\infty n]\,i\sigma^{i+}\gamma_5[\infty n + z_T\,;\,z]\, q_{i}(z)\,|P\rangle,\label{eq:Correlator}
\end{eqnarray}
 where $[x\,;\, y]$ denotes a gauge link operator 
connecting the two locations $x$ and $y$ and 
the light-like vector $n^{\mu}=(1,0,0)$. Possible complications
with slightly off-light cone vectors as suggested in TMD factorization
theorems \cite{Collins:1989gx,Ji:2004wu} are discussed below. 
Throughout this analysis we work in a covariant gauge where the transverse
gauge link at light-cone infinity is negligible. The gauge link in
(\ref{eq:Correlator}) is interpreted physically as FSIs of the active
quark with the target remnants \cite{Brodsky:2002cx,Collins:2002kn}
and is necessary for {}``naive'' time-reversal odd TMDs
\cite{Sivers:1989cc,Sivers:1990fh,Boer:1997nt} 
to exist~\cite{Collins:2002kn}.
The Boer-Mulders function appears in the factorized description of
semi-inclusive processes such as SIDIS or Drell-Yan~\cite{Mulders:1995dh,Boer:1997nt,Boer:2003cm,Bacchetta:2006tn,Ji:2004wu,Ji:2004xq,Collins:2004nx,Tangerman:1994bb,Boer:1999mm,Arnold:2008kf}
in terms of the first $k_{T}$-moment, 
$2m_{\pi}^{2}h_{1}^{\perp(1)}(x)=\int d^{2}k_{T}\,\vec{k}_{T}^{2}\, h_{1}^{\perp}(x,\vec{k}_{T}^{2})$.
It was shown in
Ref.~\cite{Boer:2003cm} that the first $k_T$-moment 
of the Boer-Mulders function
can be written in terms of a gluonic pole matrix element. 
Transforming the pion states in Eq.~(\ref{eq:Correlator}) into a mixed coordinate-momentum representation 
~\cite{Burkardt:2003uw,Meissner:2007rx} results in an impact 
parameter representation
for the gluonic pole matrix element,
\begin{eqnarray} 
\langle k_{T}\rangle(x)  \hsp&=\hsp& m_{\pi}h_{1}^{\perp(1)}(x)=\int d^2b_T\,\frac{dz^{-}}{4(2\pi)}\,\mathrm{e}^{ixP^{+}z^{-}}
\nn &&\hspace{-1cm}
\times\langle P^+,\,\vec{0}_T|\,\bar{q}(z_1)\,[z_1\,;\,z_2]\, I^{i}(z_2)\, \sigma^{i+}\, q(z_2)\,|P^+,\, \vec{0}_T\rangle, \label{eq:RelMI}
\end{eqnarray}
where the impact parameter $b_{T}$ is hidden in the arguments of
the quark fields, $z_{1/2}^\mu=\mp\frac{z^{-}}{2}n^\mu+b_T^\mu$ and 
the 4-vector   $b_T^\mu=(0,0,b_T^1,b_T^2)$.
 The operator
$I^{i}$ originates from the time-reversal behavior of the ISIs/FSIs
implemented by the gauge link operator in (\ref{eq:Correlator}) and
is given in terms of the gluonic field strength tensor $F^{\mu\nu}$,
\begin{eqnarray}
2I^{i}(z_2)=\int dy^{-}\,[z_2\,;\, y]\, gF^{+i}(y)\,[y\,;\,z_2],
\end{eqnarray}
with $y^\mu=y^- n^\mu+b_T^\mu$.

Turning our attention to GPDs of a pion, they 
are represented by an off-diagonal matrix element of a
quark-quark operator defined on the light-cone
~\cite{Diehl:2003ny,Goeke:2001tz,Belitsky:2005qn}, 
where "in"- and "out"-pion states are labeled by different
incoming and outgoing pion momenta $p$ and $p^\prime$. 
One encounters two leading twist GPDs for a pion, a chirally-even GPD $F_1^\pi$ and
the chiral odd 
 GPDs $H_{1}^{\pi}$~\cite{Meissner:2008ay}.  
We use the symmetric conventions for the kinematics for GPDs \cite{Diehl:2003ny},
$P=\frac{1}{2}(p+p^{\prime})$ and $\Delta=p^{\prime}-p$. The skewness
parameter $\xi$ is defined by $\Delta^{+}=-2\xi P^{+}$, and $t=\Delta^{2}$.
The impact parameter GPDs are obtained from the ordinary GPDs via
a Fourier-transform of the transverse momentum transfer $\vec{\Delta}_{T}$
at zero skewness $\xi=0$. The chirally-odd impact parameter GPD $\mathcal{H}_{1}^{\pi}$
 is expressed as
\begin{eqnarray}
&& \hspace{-1cm}
\int\frac{dz^{-}}{2(2\pi)}\mathrm{e}^{ixP^{+}z^{-}}\langle P^{+},\vec{0}_{T}|\,\bar{q}(z_{1})[z_{1};z_{2}]\sigma^{+i}q(z_{2})\,|P^{+},\vec{0}_{T}\rangle
\nn && \hspace{2cm}=\frac{2 b_{T}^{i}}{m_{\pi}}\,\frac{\partial}{\partial\vec{b}_{T}^{2}}\mathcal{H}_{1}^{\pi}(x,\vec{b}_{T}^{2}).\label{eq:IGPD}
\end{eqnarray}
$\mathcal{H}_{1}^{\pi}$ describes how transversely polarized
quarks are distributed in a plane transverse
to the direction of motion. 
This distribution function represents a transverse space 
distortion due to 
spin-orbit correlations~\cite{Burkardt:2005hp,Diehl:2005jf,Brommel:2007xd}.
A comparison of  the
first moment of the Boer Mulders function (\ref{eq:RelMI}) and the
first derivative of the impact parameter 
GPD $\mathcal{H}_{1}^{\pi}$,  Eq.~(\ref{eq:IGPD}),
reveals that they differ 
by  the operator $I^{i}$ which represents the FSIs. 
In various model calculations 
~\cite{Burkardt:2003je,Burkardt:2003uw,
Lu:2006kt,Meissner:2007rx}  
the FSIs 
are treated such
that the two effects of a distortion of the transverse space
parton distribution and the FSIs factorize
resulting in the  relation
\begin{eqnarray}
2m_{\pi}^2 h_{1}^{\perp(1)}(x)\simeq\int d^{2}b_{T}\,\vec{b}_T\cdot \vec{\mathcal{I}}(x,\vec{b}_{T})\,\frac{\partial}{\partial\vec{b}_{T}^{2}}\mathcal{H}_{1}^{\pi}(x,\vec{b}_{T}^{2}),\label{eq:Relation}
\end{eqnarray}
where $\mathcal{I}$ is called the 
``quantum chromodynamic 
lensing function''~\cite{Burkardt:2003uw}.
This  factorization (\ref{eq:Relation}) doesn't hold
in general~\cite{Meissner:2008ay,Meissner:2009ww}.  
On the other hand it is unknown how well Eq.~(\ref{eq:Relation})
works as a quantitative and possibly phenomenological approximation. 
A phenomenological test of Eq.~(\ref{eq:Relation})
 requires information on the parton distributions 
$h_1^{\perp (1)}$ and $H_1^\pi$
(in principle measurable)
and  quantitative  knowledge of the lensing function.
In the following sections we estimate 
the size of the lensing function using
non-perturbative eikonal methods~\cite{Abarbanel:1969ek,Fried:2000hj}
to calculate higher-order soft gluon contributions from the gauge
link and study how these soft gluons impact Eq.~(\ref{eq:Relation}).  
Up till now the relation (\ref{eq:Relation}) was used
to predict the sign of T-odd TMDs in conjunction with numbers
for the $u$- and $d$-quark contributions to the anomalous magnetic
moment of the nucleon and the assumption that FSIs
are attractive~\cite{Burkardt:2005hp}. We will also investigate
the latter assumption.

\section{TMD-GPD Relation for a Pion\label{sec3}}
We focus our attention 
on a pion in a valence
quark configuration  that  one expects for relatively large Bjorken $x$. 
Working in the spectator framework~\cite{Meyer:1990fr,Jakob:1997wg,
Ji:2002aa,Goldstein:2002vv,Gamberg:2003ey} 
and inserting a complete set of states, $\mathds{1}=\sum_{x}|X\rangle\langle X|$ in the quark correlation function Eq.~(\ref{eq:Correlator}), 
we truncate this sum 
to an antiquark and neglect multi-particle intermediate states.  
The usefulness of this approach is twofold: 
First, we are able to improve on the one gluon exchange approximation
 for  FSIs to studying 
T-odd PDFs by including higher order gluonic 
contributions and color degrees of freedom.
Second 
we are able to explore to what extent
transverse polarization effects due to T-odd PDFs  
can be described in
terms of factorization of FSIs and a spatial distortion
of impact parameter space including higher gluonic corrections~\cite{Meissner:2007rx,Meissner:2008ay} with color.
Thus, we express the pion Boer-Mulders 
function (\ref{eq:Correlator}) in the following way 
\begin{eqnarray}  
\epsilon_T^{ij} k_T^j h_1^{\perp}(x,\vec{k}_{T}^2)  = \frac{m_\pi}{8(2\pi)^3(1-x)P^+}\sum_{\sigma,d} \bar{W}i\sigma^{i+}\gamma_5 W,\label{eq:PhiSpectator}
\end{eqnarray} with the matrix element $W$ given by
\begin{eqnarray}
 W_i^{\alpha, \delta}(P,k;\sigma)=\langle P-k,\sigma,\delta|\,[\infty n\,;\,0]^{\alpha \beta}\, q_{i}^{\beta}(0)\,|P\rangle.\label{eq:W}
\end{eqnarray}
where $\sigma$ and $\delta$ represent the helicity and color of the intermediate spectator antiquark.
\begin{figure}
\centering
\includegraphics[scale=0.65]{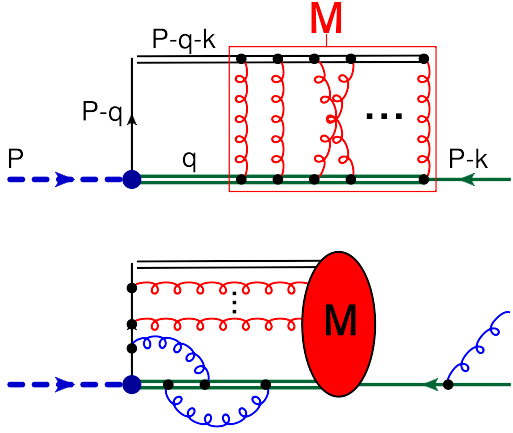}
\caption{\scriptsize The amplitude $W$ including FSIs between re-scattered
eikonalized quark and antiquark. The FSIs
are described by a non-perturbative scattering amplitude $\rM$ that is 
calculated in a generalized ladder approximation.
Gluon interactions as shown in the second  diagram are not taken into account (see text).
\label{fig:TMD-amplitude-including}}
\end{figure}
We model (\ref{eq:W}) by the diagram shown in Fig.~\ref{fig:TMD-amplitude-including}, 
where the FSIs \--- generated by the gauge link in (\ref{eq:W})
\--- are described by a non-perturbative amputated scattering amplitude
$({\rM})_{\gamma \delta}^{\alpha\beta}$ with $\beta,\,\alpha$
($\gamma,\,\delta$) color indices of incoming and outgoing quark
(antiquark). In the next section we calculate
 the scattering amplitude using 
 non-perturbative eikonal methods
thereby considering a subclass of 
possible diagrams with interactions
between quark and antiquark. 
We neglect classes of gluon exchanges in the second diagram in 
Fig.~\ref{fig:TMD-amplitude-including} represented by the red rungs
since they would be attributed to the ``interaction'' between the
quark fields and the operator $I$ in (\ref{eq:RelMI}) 
and lead to terms which break the relation (\ref{eq:Relation}). We also
neglect  real gluon emission and (self)-interactions
of quark and antiquark lines the second diagram in 
Fig.~\ref{fig:TMD-amplitude-including} 
since they represent radiative corrections of the GPD and are effectively modeled in
terms of spectator masses and a phenomenological vertex function.

The pion-quark vertex is modeled with  the interaction 
Lagrangian
\begin{eqnarray}
\mathcal{L}=-\frac{g_{\pi}}{\sqrt{N_c}} \delta^{\alpha \beta}\bar{q}^{\alpha}\gamma_{5}\vec{\tau}\cdot\vec{\varphi}q^{\beta},
\end{eqnarray}
where we allow the coupling  $g_{\pi}$
to depend on the momentum of the active quark in order to take into account
 the compositeness of the hadron and to suppress large quark virtualities~\cite{Jakob:1997wg,Gamberg:2007wm,Bacchetta:2008af}.
Applying the Feynman rules we obtain an  expression
for the matrix element $W$ in (\ref{eq:W}) from the first diagram in 
Fig.~\ref{fig:TMD-amplitude-including}
\begin{eqnarray}
\hspace{-0.725cm}W_{i,\sigma}^{\alpha\beta}(P,k) \hsp&=\hsp&\frac{-i\tau}{\sqrt{N_c}}\Bigg[\delta^{\alpha\beta}g_{\pi}(k^{2})\frac{\left[(\slash k\hspp+\hspp m_{q})v(P_s,\sigma)\right]_{i}}{k^{2}\hspp-\hspp m_{q}^{2}\hspp +\hspp i0}\hspp +\hspp\int\frac{d^{4}q}{(2\pi)^{4}}\nn
 && \hspace{-1.75cm}\frac{ g_{\pi}\left((P\hspp-\hspp q)^{2}\right)
\left[(\slash P\hspp-\hspp\slash q\hspp+\hspp m_{q})\gamma_{5}(\slash q\hspp-\hspp m_{s})
\left({\rM}\right)_{\delta\beta}^{\alpha\delta}(q,P_s)v(P_s,\sigma)\right]_{i}}{\left[n\cdot(P_s\hspp -\hspp q)+i0\right]\left[(P\hspp-\hspp q)^{2}\hspp-\hspp m_{q}^{2}\hspp +\hspp i0\right]\left[q^{2}\hspp-\hspp m_{s}^{2}\hspp+\hspp i0\right]}\Bigg],
\label{eq:StartW}
\end{eqnarray}
where $P_s\equiv P-k$ is the spectator momentum.
The first term in (\ref{eq:StartW}) represents 
the  contribution without
FSIs while the second term corresponds
to the first diagram in Fig.~\ref{fig:TMD-amplitude-including}.
We then express the FSIs through the amputated quark - antiquark scattering
amplitude ${\rM}$.  Here \emph{both} incoming quark and antiquark
are subject to the 
eikonal approximation (see, e.g.~\cite{Fried:1990ei} and references therein).
While the active quark undergoes a natural eikonalization
for a massless fermion since it represents the gauge link contribution,
the eikonalization for a massive spectator fermion is a simplification
that is justified by the physical picture of partons in an
infinite momentum frame. The eikonalization of a massive
fermion can be traced back to the Nordsieck-Bloch 
approximation~\cite{Bloch:1937pw} which describes a highly energetic
helicity conserving fermion undergoing multiple scattering
with very small momentum transfer. 
In this approximation 
the Dirac vertex 
structure $\bar{u}(p_1)\gamma^\mu u(p_2)\sim p_\mu/m\equiv v_\mu$ where
$(p_1+p_2)/2\equiv p$. For a massive anti-fermion
one identifies the velocity $v^\mu=-p^{\mu}/m$, 
and the numerator of  a fermion propagator becomes  $i(-\slash p+m)\rightarrow i(-v \cdot p+m)$.

We proceed by performing a contour-integration of the light-cone
 loop-momentum $q^{-}$ in Eq.~(\ref{eq:StartW}) where
we  consider poles which originate from the 
denominators in (\ref{eq:StartW}).
This assumes that the scattering amplitude ${\rM}$ does not
contain poles in $q^{-}$ and the integrand is well behaved on the contour in $q^-$.  
Before we proceed, it is important to point out that
one-loop calculations of T-odd functions were performed in a 
scalar diquark model~\cite{Collins:2002kn, Ji:2002aa, Gamberg:2003ey}
 and a quark target model~\cite{Goeke:2006ef} where
there are no contributions from a pole in  $q^-$ in the exchanged gluon 
propagator.
This is one reason why a factorization of the form (\ref{eq:Relation})
is exact in the one gluon exchange approximation. 
One does not expect this feature to hold in higher order calculations.
In fact  when including axial-vector diquarks~\cite{Gamberg:2007wm}
in the one-gluon exchange approximation a $q^-$ 
pole contribution from the exchanged gluon exists
which  leads to light-cone divergences when $q^- \rightarrow \infty$ 
and $q^+\rightarrow 0$. An introduction of a slightly off-light-like 
vector $\upsilon$ regulates this divergence resulting in 
logarithmic dependence of the form, 
$\rm {log}(\upsilon^+/\upsilon^-)$~\cite{Gamberg:2006ru}. 
Such logarithmic terms prevent
a factorization of the form (\ref{eq:Relation}). 
Alternatively, one may introduce 
certain vertex function $g_\pi(k^2)$ by hand that suppress contributions 
from $q^-$ - poles~\cite{Gamberg:2007wm}. Performing the contour integration on $q^-$ under these  assumptions fixes the  momentum $q^{-}$ of the antiquark
 in the loop in (\ref{eq:StartW})
to  $q^{-}=(\vec{q}_{T}^{2}+m_{s}^{2})/2q^{+}$.
  
The eikonal propagator can be split into a real and imaginary part 
using  $1/(x+i0)=\mathcal{P}(1/x)-i\pi\delta(x)$.
It has been argued in~\cite{Meissner:2007rx} that only the imaginary
part contributes to the relation (\ref{eq:Relation})
as it forces the antiquark momentum $q$ to be on the mass shell.
Thus, the imaginary part of the eikonal propagator corresponds
to a cut of the first diagram in Fig.~\ref{fig:TMD-amplitude-including}.
From the point of view of FSIs, the kinematical
point $q^{+}=(1-x)P^{+}$ is the 'natural' choice for the plus component
of the spectator. In  the  picture where one imagines the scattered
quark and antiquark  
to move quasi-collinearly with respect to the target pion
\--- backwards and forwards respectively \--- the quark and antiquark
 exchange soft gluons. Under these kinematic condition
one would expect the FSIs to be dominated by
 the {}``small'' transverse momenta of 
quark and antiquark rather than the {}``large''
plus momenta. An integration over $q^{+}$ in (\ref{eq:Wi2}) where
contributions other than the pole term contribute
include configurations where {\em large} momentum is also transferred
from quark to  antiquark in the plus direction. 
Nevertheless the principle value does contribute 
to the integral  (\ref{eq:Wi2}) which allows for
such momentum configurations. 
While this effect is beyond the picture of 
FSIs from soft gluon exchange, 
we will consider this in a future publication.  
Proceeding with the picture of 
soft gluon exchange 
there is a  clean separation of  FSIs and
spatial distortion of the parton distribution in the transverse plane
in the sense of (\ref{eq:Relation}). 
Using only the imaginary part of the eikonal propagator 
Eq.~(\ref{eq:StartW})
reduces to
\begin{eqnarray}
\hspace{-0.5cm}W_{i,\sigma}^{\alpha\beta}(P,k) \hsp&=\hsp& \frac{i\tau}{\sqrt{N_c}}(1-x)\Bigg[\delta^{\alpha\beta}g_{\pi}(k^{2})\frac{\left[(\slash k+m_{q})\gamma_{5}v(P_s,\sigma)\right]_{i}}{\vec{k}_{T}^{2}+\tilde{m}^{2}}+
\nonumber \\
 &  & \hspace{-2cm}\int\frac{d^{2}q_{T}}{(2\pi)^{2}}
\frac{g_{\pi}\left((P\hspp-\hspp q)^{2}\right)\left[(\slash P\hspp-\hspp\slash q+m_{q})
\gamma_{5}v(P_s,\sigma)\right]_{i}\left(\bar{{\rM}}\right)_{\delta\beta}^{\alpha\delta}(q;P_s)}{\left[\vec{q}_{T}^{2}+\tilde{m}^{2}\right]}\Bigg]
.\nn\label{eq:Wi2}
\end{eqnarray}
We have introduced the
notation $\bar{{\rM}}=m_{s}\rM/(2(1-x)P^{+})$.

Now we use (\ref{eq:Wi2}) to calculate the pion Boer-Mulders function
via (\ref{eq:PhiSpectator}).
Specifying the pion-quark-antiquark
vertex function 
\begin{eqnarray}
g_{\pi}(k^{2})=g_{\pi}\frac{(-\Lambda^{2})^{n-1}}{(n-1)!}\partial_{\Lambda^{2}}^{n-1}\frac{(k^{2}-m_{q}^{2})f(k^{2})}{k^{2}-\Lambda^{2}+i0},\label{eq:Vertex}\end{eqnarray}
where $f$ is a homogeneous function of the quark virtuality, 
we choose it to be a Gaussian $\exp[-\lambda^{2}|k^{2}|]$ 
in accordance with Ref.~\cite{Gamberg:2007wm}.
Inserting (\ref{eq:Wi2}) into (\ref{eq:PhiSpectator}) and a bit of algebra
yields the following expression for the Boer-Mulders function
\begin{eqnarray}
\hspace{-.6cm}
\epsilon_{T}^{ij}k_{T}^{j}h_{1}^{\perp}(x,\vec{k}_{T}^{2})\hsp& = \hsp&\frac{2g_{\pi}^{2}m_{\pi}}{(2\pi)^{3}\Lambda^2}(xm_{s}+(1-x)m_{q})\left((1-x)\Lambda^{2}\right)^{2n-1}
\nonumber \\
 &  & \hspace{-2.0cm}\times\int\frac{d^{2}q_{T}}{(2\pi)^{2}}\frac{d^{2}p_{T}}{(2\pi)^{2}}\,\epsilon_{T}^{ji}(q_{T}^{j}-p_{T}^{j})\frac{\mathrm{e}^{-\frac{2\lambda^{2}}{1-x}(xm_{s}^{2}-x(1-x)m_{\pi}^{2})}
\mathrm{e^{-\frac{\lambda^{2}}{1-x}(\vec{q}_{T}^{2}+\vec{p}_{T}^{2})}}}{\left[\vec{q}_{T}^{2}+\tilde{\Lambda}^{2}(x)\right]^{n}\left[\vec{p}_{T}^{2}+\tilde{\Lambda}^{2}(x)\right]^{n}}\nn
 & & \hspace{-2.0cm}\times\left(\Im[{\bar{\rM}}^{\mathrm{eik}}]\right)_{\delta\beta}^{\alpha\delta}(\vec{k}_{T}+\vec{q}_{T})\Big((2\pi)^{2}\delta^{\alpha\beta}\delta^{(2)}(\vec{p}_{T}+\vec{k}_{T})
\nn &&\hspace{2.0cm}+\left(\Re[\bar{{\rM}}^{\mathrm{eik}}]\right)_{\gamma\alpha}^{\beta\gamma}(\vec{k}_{T}+\vec{p}_{T})\Big),
\label{eq:BMNo}
\end{eqnarray}
with $\tilde{\Lambda}^{2}(x)=xm_{s}^{2}-x(1-x)m_\pi^{2}+(1-x)\Lambda^{2}$.
Anticipating an eikonal form for the scattering amplitude
$\bar{{\rM}}(x,\vec{k}_{T},\vec{q}_{T})\rightarrow\bar{{\rM}}^{\mathrm{eik}}(|\vec{q}_{T}+\vec{k}_{T}|)$
that will be discussed in the next section
we exploit this property to simplify the expression
and show a relation to the chirally-odd GPD $H_{1}^{\pi}$. 
Since GPDs are defined from 
collinear light-cone correlations functions gauge link contributions to GPDs don't lead to an observable effect. In fact, in light-cone gauge the corresponding contributions from the gauge link are re-shuffled into
the gluon propagators~\cite{Ji:2002aa} and they appear as 
gluon dressings of the tree-level contribution to GPDs.
 Thus one can consistently describe GPDs
 from tree-level diagrams in the spectator model where the
effects of gluon dressings are effectively hidden in the model parameters.
A calculation for the GPD $H_{1}^{\pi}$  
for an antiquark spectator can be found in~\cite{Meissner:2008ay}.
It is easy to generalize it with a phenomenological vertex function
(\ref{eq:Vertex}). We obtain the following representation 
\begin{eqnarray}
\hspace{-.6cm}H_{1}^{\pi}(x,0,-\vec{\Delta}_{T}^{2}) \hsp&=\hsp& \frac{-g_{\pi}^{2}m_{\pi}}{2(2\pi)^{3}\Lambda^{2}}(xm_{s}\hspp+(1\hspp-\hspp x)\ m_{q})
\left(\hspp\frac{(1\hspp-\hspp x)\Lambda^{2}}{\vec{D}_{T}^{2}\hspp+\hspp \tilde{\Lambda}^{2}(x)}\hspp\right)^{2n-1}
\nn &  & \hspace{-2cm}\times\int_{0}^{2\pi}d\varphi\int_{0}^{1}dz\frac{z^{2n-2}\mathrm{e}^{2\lambda^{2}\Lambda^{2}}\mathrm{e}^{-\frac{2\lambda^{2}(\vec{D}_{T}^{2}
+\tilde{\Lambda}^{2}(x))}{(1-x)z}}}{\left[1-4z(1-z)\frac{\vec{D}_{T}^{2}}{\vec{D}_{T}^{2}+
\tilde{\Lambda}^{2}(x)}\cos^{2}\varphi\right]^{n}},
\label{eq:H1Pi} 
\end{eqnarray}
where $\vec{D}_{T}^{2}=\frac{1}{4}(1-x)^{2}\vec{\Delta}_{T}^{2}$.  
Performing a translation of the integration variables in (\ref{eq:BMNo})
according to $q_{T}\rightarrow q_{T}+k_{T}$ and $p_{T}\rightarrow p_{T}+k_{T}$,
a rotation of the form $q_T^\prime=q_{T}-p_{T}$, $p_T^\prime=q_{T}+p_{T}$, weighting with a transverse quark vector $k_{T}^{i}$ and integrating both sides
over $k_{T}$ we  find the relation 
\begin{eqnarray}
\hspace{-0.7cm}m_{\pi}^{2}h_{1}^{\perp(1)}(x)\hsp&=\hsp &\int\frac{d^{2}q_{T}}{2(2\pi)^{2}}\,\vec{q}_{T}\cdot\vec{I}(x,\vec{q}_{T})H_{1}^{\pi}\left(x,0,-\left(\frac{\vec{q}_{T}}{1-x}\right)^{2}\right)\,.
\label{eq:RelationDiquark}
\end{eqnarray}
The function $I^{i}$ can be expressed in terms of the real
and imaginary part of the scattering amplitude $\bar{{\rM}}$,
\begin{eqnarray}
\hspace{-0.5cm}I^{i}(x,\vec{q}_{T})\hsp&=\hsp&
\frac{1}{N_{c}}\int\frac{d^{2}p_{T}}{(2\pi)^{2}}\,(2p_{T}-q_{T})^{i}\,
\left(\Im[\bar{{\rM}}^{\mathrm{eik}}]\right)_{\delta\beta}^{\alpha\delta}
(|\vec{p}_{T}|)
\nn &&
\hspace{-1cm}\Big((2\pi)^{2}\delta^{\alpha\beta}\delta^{(2)}(\vec{p}_{T}-\vec{q}_{T})
+\left(\Re[\bar{{\rM}}^{\mathrm{eik}}]\Big)_{\gamma\alpha}^{\beta\gamma}(|\vec{p}_{T}-\vec{q}_{T}|)\right).
\label{eq:LensFunc}
\end{eqnarray}
In order to derive the relation (\ref{eq:Relation}) one transforms
Eq.~(\ref{eq:RelationDiquark}) into the impact parameter space
via a Fourier transforms of the following form,
\begin{eqnarray}
\mathcal{H}_{1}^{\pi}(x,\vec{b}_{T}^{2})=\int\frac{d^{2}\Delta_{T}}{(2\pi)^{2}}\,\mathrm{e}^{-i\vec{\Delta}_{T}\cdot\vec{b}_{T}}H_{1}^{\pi}(x,0,-\vec{\Delta}_{T}^{2}).
\end{eqnarray}
The lensing function in the impact parameter space then reads,
\begin{eqnarray}
\mathcal{I}^{i}(x,\vec{b}_{T}) \hsp&=\hsp&
 i(1-x)\int\frac{d^{2}q_{T}}{(2\pi)^{2}}\,\mathrm{e}^{i\frac{\vec{q}_{T}\cdot\vec{b}_{T}}{1-x}}I^{i}(x,\vec{q}_{T}).\label{eq:LensingFunctioneik}
\end{eqnarray}
In the following section we will use a quark-antiquark scattering
amplitude computed in relativistic eikonal models as input for the
lensing function (\ref{eq:LensFunc}).

\section{The Lensing Function in Relativistic Eikonal Model}

In order to calculate the  2 $\rightarrow$ 2 scattering amplitude $\rM$ 
(needed for (\ref{eq:LensFunc}))
we use functional methods to incorporate the color degrees of freedom 
in the eikonal limit when soft gauge bosons couple to highly
energetic matter particles on the light cone.
It is non-trivial to extend the  functional methods established
in an Abelian to  non-Abelian gauge theory such as QCD.
Attempts in this
direction were made in Refs.~\cite{Fried:1996uv,Fried:2000hj}, and
only recently a fully Lorentz and gauge invariant treatment was presented
in Ref.~\cite{Fried:2009fw}. 
Here we outline  the details of the functional approach
as it pertains to implementing color structure to the scattering
amplitude $\rM$ and thereby the lensing function. We leave the
details to a forthcoming publication~\cite{gam_schle_sivers}.

Starting from the generating functional $Z$ for QCD in a covariant 
gauge, a quark antiquark 4-point
function $T$ can then be defined by functional derivatives with respect
to quark sources which yields,
\begin{eqnarray}
\hspace{-0.7cm}T_{2\rightarrow 2}&\hsp\propto&\hsp\int\mathcal{D}A\,\mathrm{e}^{-\frac{i}{4}\int(F^{2}+2\lambda(\partial\cdot A)^{2})}\,\mathrm{e}^{\Tr\ln G^{-1}[A]+\Tr\ln H^{-1}[A]}\, G[A]\,\bar{G}[A].\nn\label{eq:4-pointstart}
\end{eqnarray}
The first exponential describes the gluonic part of the theory
including self-interactions and the second exponential describes internal
closed quark and ghost loops. $G$, $\bar{G}$ are the non-perturbative
quark- and antiquark-propagator determining the external legs of the
4-point function $T$, and  $H$ is the ghost propagator~\cite{gam_schle_sivers}. 
 One imposes  eikonal approximations
on these propagators~\cite{Abarbanel:1969ek,Fried:2000hj}
 that simplify the computation of the path-integral.
In an Abelian theory the  eikonal approximation
as discussed in the previous section leads to a well-known eikonal
representation~\cite{Abarbanel:1969ek}, which
was argued in~\cite{Fried:1996uv,Fried:2000hj} to generalize to QCD
in the following way, e.g. for a massless fermion
\begin{eqnarray}
\hspace{-0.7cm}G_{\alpha\beta}^{\mathrm{eik}}(x,y|A)&\hsp=&\hsp-i\int_{0}^{\infty}ds\delta^{(4)}(x-y-sn)\left(\mathrm{e}^{-ig\int_{0}^{s}d\beta\, n\cdot A^{a}(y+\beta n)t^{a}}\right)_{\alpha\beta}^{+},\label{eq:eikprop}\nn\end{eqnarray}
where color is implemented by a path-ordered exponential indicated
by the brackets $(...)^{+}$ and the color matrix $t^{a}$ in the
exponential. 

Inserting the eikonal representation for the quark- and antiquark
propagator into Eq.~(\ref{eq:4-pointstart}) and  implementing the generalized ladder approximation 
one finds the {\em color gauge invariant}
result corresponding to the 
picture of FSIs  discussed in the previous section,
\begin{eqnarray}
\hsp\left({\rM}^{\mathrm{eik}}\right)_{\delta\beta}^{\alpha\delta}(x,|\vec{q}_{T}+\vec{k}_{T}|) \hsp&=\hsp&\frac{(1-x)P^{+}}{m_{s}}\int d^{2}z_{T}\,\mathrm{e}^{-i\vec{z}_{T}\cdot(\vec{q}_{T}+\vec{k}_{T})}\label{eq:eikonalAmplitude}\\
 &  & \hspace{-3.5cm}\times\Bigg[\int d^{N_{c}^{2}-1}\alpha\int\frac{d^{N_{c}^{2}-1}u}{(2\pi)^{N_{c}^{2}-1}}\,\mathrm{e}^{-i\alpha\cdot u}\left(\mathrm{e}^{i\chi(|\vec{z}_{T}|)t\cdot\alpha}\right)_{\alpha\delta}\left(\mathrm{e}^{it\cdot u}\right)_{\delta\beta}-\delta_{\alpha\beta}\Bigg].\nonumber \end{eqnarray}
In this expression, the ($N_{c}^{2}-1$) dimensional integrals result
from  auxiliary fields $\alpha^{a}(s)$
and $u^{a}(s)$ that  were introduced in the functional formalism (see
Ref.~\cite{Fried:2000hj})  to separate the physical gluon
fields from the color matrices.  The eikonal phase $\chi(|\vec{z}_{T}|)$ in Eq.~(\ref{eq:eikonalAmplitude})
represents the arbitrary amount of soft gluon exchanges that are summed
up into an exponential form and is expressed in terms of the gluon
propagator in a covariant gauge,
\begin{eqnarray}
\chi(|\vec{z}_{T}|)=g^{2}\int_{-\infty}^{\infty}d\alpha\int_{-\infty}^{\infty}d\beta\, n^{\mu}\bar{n}^{\nu}\mathcal{D}_{\mu\nu}(z+\alpha n-\beta\bar{n}),\label{eq:PhaseEik}
\end{eqnarray}
where $\mathcal{D}$ denotes the gluon propagator, and $g$ is 
the strong coupling. In this form the  4-velocity vector $v^\mu$
is expressed in terms of the 
complementary light cone vector $\bar{n}$ where 
$v=-\frac{(1-x)P^{+}}{m_{s}}\bar{n}$,
with $n\cdot\bar{n}=1$ and $\bar{n}^{2}=0$. 
One may choose $\bar{n}=(0,1,\vec{0}_{T})$.

In Eq.~(\ref{eq:eikonalAmplitude})
we evaluate the color integral,
\begin{eqnarray}
\hspace{-0.7cm}f_{\alpha\beta}(\chi)&\hsp\equiv\hsp&\hsp\int\hspp 
d^{N_{c}^{2}-1}\alpha\hspp\int\frac{d^{N_{c}^{2}-1}u}{(2\pi)^{N_{c}^{2}-1}}\,\mathrm{e}^{-i\alpha\cdot u}\left(\mathrm{e}^{i\chi(|\vec{z}_{T}|)t\cdot\alpha}\right)_{\alpha\delta}\left(\mathrm{e}^{it\cdot u}\right)_{\delta\beta}\hspp-\hspp\delta_{\alpha\beta}\nn
\label{eq:ColorIntegral}
\end{eqnarray}
by deriving a power series representation for arbitrary $N_c$.
We  expand the exponential $\exp[i\chi t\cdot\alpha]$ and rewrite
the resulting factors as derivatives with respect to $u$. Then we
perform integrations by parts which reduces the $\alpha$ integral
to a simple $\delta$-function. This simplifies the $u$-integral 
where $u$ is set to zero after differentiation 
We obtain 
\begin{eqnarray}
f_{\alpha\beta}(\chi)=\sum_{n=1}^{\infty}\frac{(i\chi)^{n}}{n!}(-i)^{n}(t^{a_{1}}...t^{a_{n}})_{\alpha\delta}\frac{\partial^{n}(\mathrm{e}^{it\cdot u})_{\delta\beta}}{\partial u^{a_{1}}...\partial u^{a_{n}}}\Big|_{u=0}.
\label{eq:Aux1}
\end{eqnarray}
Now we  expand the remaining exponential in Eq.~(\ref{eq:Aux1})
and note that one can write the set of partial derivatives with respect
to $u^{a_i}$  as a sum over all
permutations $P_{n}$ of the set $\{1,...,n\}$, which results in the  power series representation for
$f$,
\begin{eqnarray}
\hspace{-0.68cm}f_{\alpha\beta}(\chi)\hsp&=&\hsp\sum_{n=1}^{\infty}\frac{(i\chi)^{n}}{(n!)^{2}}\sum_{a_{1}=1}^{N_{c}^{2}-1}...\sum_{a_{n}=1}^{N_{c}^{2}-1}\sum_{P_{n}}\left(t^{a_{1}}...t^{a_{n}}t^{a_{P_{n}(1)}}...t^{a_{P_{n}(n)}}\right)_{\alpha\beta}.\label{eq:PowerSeries}\end{eqnarray}
This color factor matrix nicely illustrates the generalized ladder
approximation. If only direct ladder gluons were considered 
the sum over permutations would become trivial in
Eq.~(\ref{eq:PowerSeries}) and only terms 
$(t^{a_{1}}...t^{a_{n}}t^{a_{n}}...t^{a_{1}})_{\alpha\beta}=C_{F}^{n}\delta_{\alpha\beta}$
with $C_{F}=\frac{N_{c}^{2}-1}{2N_{c}}$ would contribute.  This constitutes the leading order in a large-$N_c$ expansion while
non-planar diagrams, i.e. crossed gluon graphs, are suppressed. For
the leading contribution one may simply replace $\alpha\rightarrow C_{F}\alpha_{s}$
and work in an Abelian theory. In particular, this replacement
was suggested in perturbative model calculations~\cite{Brodsky:2002cx,Brodsky:2002rv}.
Since we take into account crossed gluons we have to 
sum over all permutations in (\ref{eq:PowerSeries}),
and such a  replacement is not possible. 
In an Abelian theory, the generating matrices $t$ 
reduce to unity, $t=1$, and since we have $n!$ permutations of the
set $\{1,...,n\}$, we recover the well-known result for the Coulomb phase,

\begin{eqnarray}
f^{U(1)}(\chi)=\sum_{n=1}^{\infty}\frac{(i\chi)^{n}}{n!}=\mathrm{e}^{i\chi}-1.\label{eq:CFU1}
\end{eqnarray}
For the non-Abelian $N_{c}=2$ theory 
the generators are given by the Pauli matrices 
$\sigma^{a}=2t^{a}$. Instead of using the power series representation
we can calculate the integral (\ref{eq:ColorIntegral}) analytically
by means of the relation $\left(\mathrm{e}^{iu\cdot\frac{\sigma}{2}}\right)_{\alpha\beta}=\delta_{\alpha\beta}\cos\left(\frac{|u|}{2}\right)+\frac{i\vec{\sigma}_{\alpha\beta}\cdot\vec{u}}{|u|}\sin\left(\frac{|u|}{2}\right)$.
We obtain a slightly different result compared to
 Ref.~\cite{Fried:2000hj} for SU(2),
\begin{eqnarray}
\hspace{-0.7cm}f_{\alpha\beta}^{SU(2)}(\frac{\chi}{4})\hsp &=&\hsp\delta_{\alpha\beta}\left(\cos\frac{\chi}{4}-\frac{\chi}{4}\sin\frac{\chi}{4}\hspp-\hspp 1+i\left(2\sin\frac{\chi}{4}\hspp+\hspp\frac{\chi}{4}\cos\frac{\chi}{4}\right)\right).\nn\label{eq:SU2Analytical}
\end{eqnarray}
As a check on our numerical and analytical
approaches we 
numerically calculate the lowest coefficients in the power
series (\ref{eq:PowerSeries}), and they agree with the coefficients
in an expansion in $\chi$ of the analytical result (\ref{eq:SU2Analytical}). 
The disadvantage of using the
power series representation (\ref{eq:PowerSeries}) 
is apparent for  numerical calculations since
 the number of operations grows with $n!$. That said, 
for SU(2) we calculated the first eight coefficients.  For QCD, $N_{c}=3$, the generators $t$ are given by the Gell-Mann
matrices  $\lambda^{a}=2t^{a}$. Due to the 
difficulty of integrating over the Haar measure in Eq.~(\ref{eq:ColorIntegral})
we put off the analytical treatment~\cite{gam_schle_sivers}.
Using the power series (\ref{eq:PowerSeries}) we
derive the following approximative color function 
for $a=\chi/4$ 
\begin{eqnarray}
\Re[f_{\alpha\beta}^{SU(3)}](a) \hsp&=\hsp&
 \delta_{\alpha\beta}(-c_{2}a^{2}+c_{4}a^{4}-c_{6}a^{6}-c_{8}a^{8}+...),\nn
\Im[f_{\alpha\beta}^{SU(3)}](a) \hsp&=\hsp& \delta_{\alpha\beta}(c_{1}a-c_{3}a^{3}+c_{5}a^{5}-c_{7}a^{7}+...),\label{eq:CFSU3Im}
\end{eqnarray}
with the numerical values $c_{1}=5.333$, $c_{2}=6.222$, $c_{3}=3.951$,
$c_{4}=1.934$, $c_{5}=0.680$, $c_{6}=0.198$, $c_{7}=0.047$, $c_{8}=0.00967$.
\begin{figure*}
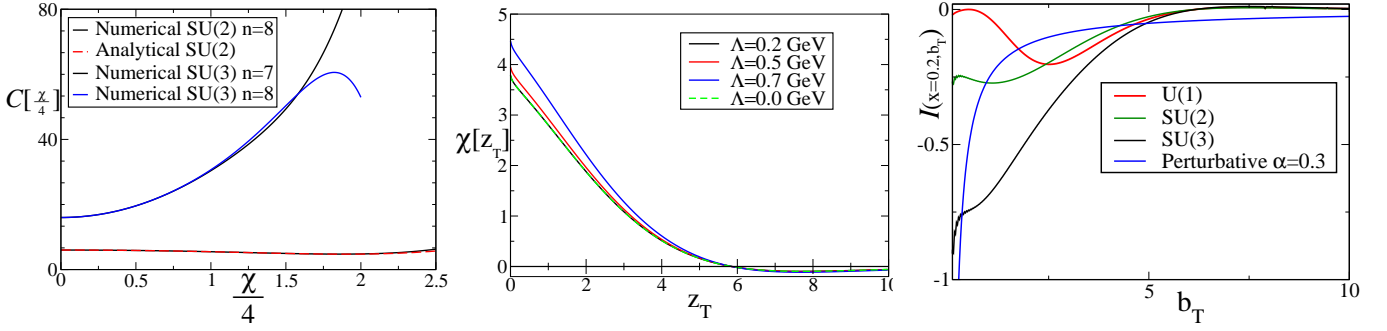

\centering
\begin{minipage}{2.3in}
\includegraphics[width=2.3in]{SU2_3Color}
\end{minipage}
\begin{minipage}{2.3in}
\includegraphics[width=2.3in]{phase}
\end{minipage}
\hspace{0.05cm}
\begin{minipage}{2.3in}
\includegraphics[width=2.3in]{LensingIP}
\end{minipage}
\caption{\scriptsize \textbf{Left:} 
$C[\frac{\chi}{4}]$ of Eq.~(\ref{eq:Cchi}) as a function
of the eikonal amplitude $\frac{\chi}{4}$. 
We compare the numerical result computed by means of Eq.~(\ref{eq:PowerSeries})
up to the order $n=8$ with the analytical result in Eq.~(\ref{eq:LensIPSU2})
for the $SU(2)$ color case. The numerical  and  
analytical result agree up to $\frac{\chi}{4}\sim2$.
For $SU(3)$, we compare the numerical results for the
orders $n\,=7,\,8$. The results are reliable for
 $\frac{\chi}{4}\sim1.5$.
\textbf{Center:}
The eikonal phase $\chi^{DS}(|\vec{z}_{T}|)$ vs. $|\vec{z}_{T}|$
with input from Dyson-Schwinger equations at scales $\Lambda_{QCD}=0\,\mathrm{GeV},\,0.2\,\mathrm{GeV},\,0.5\,\mathrm{GeV},\,0.7\,\mathrm{GeV}$.
\textbf{Right:}
The lensing function $\mathcal{I}^{i}(x,\vec{b}_{T})$
from Eq.~(\ref{eq:LensColor}) for U(1), SU(2) and SU(3)
for $x=0.2$ at a scale $\Lambda_{QCD}=0.2\,\mathrm{GeV}$. For comparison
we also plot the perturbative result of Ref.~\cite{Meissner:2008ay}
including the eikonalized antiquark spectator
with an arbitrary value for the coupling,  $\alpha=0.3$.
\label{fig:The-function-}}
\end{figure*}
Working in coordinate space 
we express the lensing function directly in terms of the eikonal
phase $\chi$ defined in Eq.~(\ref{eq:PhaseEik}). Defining the 
eikonal amplitude 
as in section~\ref{sec3},  
where the real and imaginary part are 
\begin{eqnarray}
\Re[\bar{{\rM}}_{\alpha\beta}](\vec{p)} & = & \frac{1}{2}\int d^{2}z\,\mathrm{e}^{i\vec{p}\cdot\vec{z}}\Re[f_{\alpha\beta}(\chi(|\vec{z}_{T}|))],\label{eq:Re}\\
\Im[\bar{{\rM}}_{\alpha\beta}](\vec{p}) & = & \frac{1}{2}\int d^{2}z\,\mathrm{e}^{i\vec{p}\cdot\vec{z}}\Im[f_{\alpha\beta}(\chi(|\vec{z}_{T}|))], \label{eq:Im}\end{eqnarray}
we insert (\ref{eq:Re}) and (\ref{eq:Im}) into the lensing function
(\ref{eq:LensFunc}) then transform it via (\ref{eq:LensingFunctioneik})
into the impact parameter space. 
This yields a lensing function of the form,
\begin{eqnarray}
\hspace{-.7cm} \mathcal{I}^{i}(x,\vec{b}_{T})&=&\frac{(1-x)}{2N_{c}}\frac{b_{T}^{i}}{|\vec{b}_{T}|}\frac{\chi^{\prime}}{4}C[\frac{\chi}{4}],\label{eq:LensColor}\nn
\hspace{-2cm}C[\frac{\chi}{4}]&\equiv &\Bigg[\left(\Tr\Im[f]\right)^{\prime}(\frac{\chi}{4})+\frac{1}{2}\Tr\left[\left(\Im[f]\right)^{\prime}(\frac{\chi}{4})\left(\Re[f]\right)(\frac{\chi}{4})\right]\nn
& & -\frac{1}{2}\Tr\left[\left(\Im[f]\right)(\frac{\chi}{4})\left(\Re[f]\right)^{\prime}(\frac{\chi}{4})\right]\Bigg],
\label{eq:Cchi}
\end{eqnarray}
where $\chi^{\prime}$ denotes the first derivative with respect to
$|\vec{z}_{T}|$, and $\left(\Im[f]\right)^{\prime}$ and $\left(\Re[f]\right)^{\prime}$ are the first derivatives of the real and imaginary parts of the color
function $f$. Also, the eikonal phase  
is understood to be a function of $|\vec{b}_{T}|/(1-x)$. 
Inserting (\ref{eq:CFU1}) into (\ref{eq:LensColor})
results into the following expression for the lensing function in
an Abelian U(1)-theory 
\begin{eqnarray}
\hspace{-0.7cm}\mathcal{I}_{U(1)}^{i}(x,\vec{b}_{T})&\hsp=&\hsp (1-x)\frac{b_{T}^{i}}{4|\vec{b}_{T}|}\chi^{\prime}(\frac{|\vec{b}_{T}|}{1-x})\left(1+\cos\chi(\frac{|\vec{b}_{T}|}{1-x})\right).\label{eq:LensIPU1}\end{eqnarray}
Similarly  from (\ref{eq:SU2Analytical}) we calculate the lensing
function in an SU(2)-theory
\begin{eqnarray}
\mathcal{I}_{SU(2)}^{i}(x,\vec{b}_{T})&\hsp=\hsp&\frac{(1-x)b_{T}^{i}}{16|\vec{b}_{T}|}\chi^{\prime}(\frac{|\vec{b}_{T}|}{1-x}) \\
 & & \hspace{-2.5cm}\times\Bigg(3(1+\cos\frac{\chi}{4})+\left(\frac{\chi}{4}\right)^{2}
-\sin\frac{\chi}{4}\left(\frac{\chi}{4}-\sin\frac{\chi}{4}\right)\Bigg)\Big|_{\chi=\chi\left(\frac{|\vec{b}_{T}|}{1-x}\right)}.\nnn\label{eq:LensIPSU2}
\end{eqnarray}
For the SU(3)-QCD case we use Eq.~(\ref{eq:CFSU3Im}).
In Fig.~\ref{fig:The-function-}
the function $C[\frac{\chi}{4}]$ is plotted versus $\frac{\chi}{4}$.
While the convergence of the power series
is slightly better for SU(2) where the
numerical result, calculated to eighth order, agrees with the
analytical result up to $\frac{\chi}{4}\sim2$, 
we can trust the numerical result computed with eight coefficients up
to $\frac{\chi}{4}\sim1.5$ for SU(3).

At this point we discuss the eikonal phase $\chi$ 
as defined in (\ref{eq:PhaseEik}) which is determined by two quantities,
the strong coupling $g$ and the gluon propagator $\mathcal{D}$.
One can write a general form for the gluon propagator in momentum
space
\begin{eqnarray}
\hspace{-0.6cm}\mathcal{D}_{\mu\nu}^{ab}(z) & \hsp=\hsp & \delta^{ab}\int\frac{d^{4}k}{(2\pi)^{4}}\,\tilde{\mathcal{D}}_{\mu\nu}(k)\mathrm{e}^{-ik\cdot z}\nonumber\\
 & \hsp\equiv\hsp &\delta^{ab}\int\frac{d^{4}k}{(2\pi)^{4}}\,\left[g_{\mu\nu}\tilde{\mathcal{D}}_{1}(k^{2})+k_{\mu}k_{\nu}\tilde{\mathcal{D}}_{2}(k^{2})\right]\mathrm{e}^{-ik\cdot z},\label{eq:GluePropParam}
\end{eqnarray}
where the gauge dependent part is  in $\tilde{\mathcal{D}}_{2}$.
However,  the gauge dependent part does not appear in the eikonal
phase when inserting Eq.~(\ref{eq:GluePropParam}) into Eq.~(\ref{eq:PhaseEik})
because the  eikonal vectors $n$ and $v\simeq-\frac{(1-x)P^{+}}{m_{s}}\bar{n}$
are light-like. Performing the integral yields the following expression
for the eikonal phase 
\begin{eqnarray}
\chi(|\vec{z}_{T}|)=\frac{g^{2}}{2\pi}\int_{0}^{\infty}dk_{T}\, k_{T}\, J_{0}(|\vec{z}_{T}|k_{T})\tilde{\mathcal{D}}_{1}(-k_{T}^{2}),
\label{eq:Phase}
\end{eqnarray}
where $J_{0}$ is a Bessel function of the first 
kind.  The gluon propagator represents all exponentiated gluons 
exchanged between the two eikonal lines in the generalized ladder
approximation in Fig.~\ref{fig:TMD-amplitude-including}. The couplings
represent the strength of the quark (antiquark) - gluon interaction
in Fig.~\ref{fig:TMD-amplitude-including}.

As a check of the calculation we investigated the perturbative limit
of our calculation. Assuming that the quark - gluon interaction $g^2$
is small and using  perturbative gluon propagator in Feynman
gauge for $\tilde{\mathcal{D}}_1$ one can expand our non-perturbative 
result in Eq.~(\ref{eq:Cchi}) to $g^2$. The leading order corresponds to
the result of the one-loop calculation of the Boer-Mulders function of Ref.~\cite{Meissner:2008ay}
after additional eikonalization of the antiquark.

\section{Non-perturbative Quantities from the Dyson-Schwinger approach}

In order to obtain a numerical estimate for the eikonal phase, 
 it is important to have a realistic estimate of the size of the QCD
coupling $g$ or $\alpha_{s}=\frac{g^{2}}{4\pi}$. 
Since all the gluons exchanges between the eikonal lines are soft,
the interactions take place at a soft scale. Thus we need to know
the running of the strong coupling in the infrared limit. 
Inserting a perturbative  gluon propagator might 
not describe the gluon exchange realistically.
One would expect that a
non-perturbative gluon propagator would be a better choice.
The infrared behavior of both quantities, the running of the strong
coupling and the non-perturbative gluon propagator, have been studied
in the framework of the Dyson-Schwinger equations~\cite{Fischer:2003rp,Alkofer:2003jj,Alkofer:2006jf,Fischer:2008uz}
 and also in lattice(see e.g.~\cite{Sternbeck:2008mv}).   One learns from such studies that the strong coupling
has a value of about $\alpha_{s}(0)\simeq2.972$ in the infrared limit.
In particular in Ref.~\cite{Fischer:2003rp} fits were presented for
the running coupling. Since we are merely interested in a numerical
estimate of the lensing function we will apply the simplest form of
the running coupling presented in \cite{Fischer:2003rp},
\begin{eqnarray}
\alpha_{s}(\mu^{2})=\frac{\alpha_{s}(0)}{\ln\left[\mathrm{e+a_{1}(\mu^{2}/\Lambda^{2})^{a_{2}}+b_{1}(\mu^{2}/\Lambda^{2})^{b_{2}}}\right]}.
\label{eq:runningalpha}
\end{eqnarray}
The values for the fit parameters are $\Lambda=0.71\,\mathrm{GeV}$,
$a_{1}=1.106$, $a_{2}=2.324$, $b_{1}=0.004$ and $b_{2}=3.169$. 
These calculations were performed in Euclidean space where
Landau gauge was applied, and agree reasonably well with each other.
Because the light cone components in Eq.~(\ref{eq:Phase}) are already
integrated out and the remaining integration range is over a 2-dimensional
transverse Euclidean space, and because the gauge dependent part of
the gluon propagator does not contribute, it is natural to apply the Euclidean
results in Landau gauge of the Dyson-Schwinger framework. One unique
feature of Dyson-Schwinger studies of the gluon propagator is that
it rises like $(k^{2})^{2\kappa-1}$ in the infra-red limit with a
universal coefficient $\kappa\simeq0.595$. This makes it infrared
finite in contrast to the perturbative propagator. 
A fit to the results for the non-perturbative gluon propagator has been given in Ref.~\cite{Fischer:2002hna,Fischer:2003rp,Alkofer:2008tt},
\begin{eqnarray}
\hspace{-0.4cm} Z(p^{2},\mu^{2})& \, =\,& p^{2}\mathcal{D}^{-1}(p^{2},\mu^{2}) \nonumber \\
 & & \hspace{-0.6cm}= \left(\frac{\alpha_{s}(p^{2})}{\alpha_{s}(\mu^{2})}\right)^{1+2\delta}\left(\frac{c\left(\frac{p^{2}}{\Lambda^{2}}\right)^{\kappa}+d\left(\frac{p^{2}}{\Lambda^{2}}\right)^{2\kappa}}{1+c\left(\frac{p^{2}}{\Lambda^{2}}\right)^{\kappa}+d\left(\frac{p^{2}}{\Lambda^{2}}\right)^{2\kappa}}\right)^{2},
\label{eq:fit}
\end{eqnarray}
with the parameters $c=1.269$, $d=2.105$, and $\delta=-\frac{9}{44}$.
These fits for the running coupling and the gluon propagator merge
with the spirit of the eikonal methods described above since closed
fermion loops (quenched approximation) were neglected. By using the
non-perturbative propagator (\ref{eq:fit}), we partly reintroduce
gluon self-interactions 
that were originally
neglected in the generalized ladder approximation. According to Ref.~\cite{Alkofer:2008tt}
the fitting functions Eqs.~(\ref{eq:runningalpha})
and (\ref{eq:fit}) were adjusted to Dyson-Schwinger results obtained
at a very large renormalization scale, the mass of the top quark,
$\mu^{2}=170\,\mathrm{GeV}^{2}$, which defines the normalization
in (\ref{eq:fit}). Since the lensing function deals with soft
physics, intuitively we prefer a much lower hadronic scale which sets
the normalization, $\mu=\Lambda_{QCD}\approx 0.2\,\mathrm{GeV}$. In
the spirit of Sudakov form factors we also assume that the scale at
which the gluons are exchanged is given by the transverse gluon momentum
that we integrate over. In this way the running coupling serves as
a vertex form factor that additional cuts off large gluon transverse
momenta.

Our  ansatz for the eikonal phase given by Dyson-Schwinger
quantities then reads,\begin{eqnarray}
\hspace{-0.68cm}\chi^{DS}(|\vec{z}_{T}|)\hsp&=&\hsp2\int_{0}^{\infty}dk_{T}\, k_{T}
\alpha_{s}(k_{T}^{2})J_0(|\vec{z}_{T}|k_T)
Z(k_{T}^{2},\Lambda_{QCD}^{2})/k_{T}^{2}.
\end{eqnarray}
The numerical result for this ansatz is shown in the center panel of  
Fig.~\ref{fig:The-function-}.
We plot this function for various scale $\Lambda_{QCD}=0\,\mathrm{GeV}$,
$0.2\,\mathrm{GeV}$, $0.5\,\mathrm{GeV}$, $0.7\,\mathrm{GeV}$.
Although the choice of this scale is rather arbitrary we observe only
a very mild dependence  on this
scale as long as it remains soft. We further observe that the phase
doesn't exceed a value of $4$ - $4.5\, \rightarrow \chi_{\mathrm max}/4\approx 1.15$. 
Thus this feature makes the application of the power series of the color function
in SU(3) reliable since $\chi/4$ never exceeds $1.5$ in the lensing function,
Eq.~(\ref{eq:Cchi}) and in turn in the calculation of the Boer-Mulders
function in Eq.~(\ref{eq:Relation}).

Finally, we insert our ansatz for the eikonal phase into the lensing
functions (\ref{eq:LensColor}) for a U(1), SU(2) and SU(3)
color function. We plot the results in Fig.~\ref{fig:The-eikonal-phase}
for a color function for U(1), SU(2), SU(3). 
While we observe
that all lensing functions fall off at large transverse distances,
they are quite different in size at small distances.   
However  for each case,  
the all order calculation sums up to  an exponential of the eikonal phase
where one observes oscillations from the Bessel function 
$J_{0}$ of the first kind. Despite these oscillations the lensing function 
remains negative. 
\section{The Pion Boer-Mulders Function}
\begin{figure}
\begin{center}
\begin{minipage}{3.0in}
\includegraphics[width=3.0in]{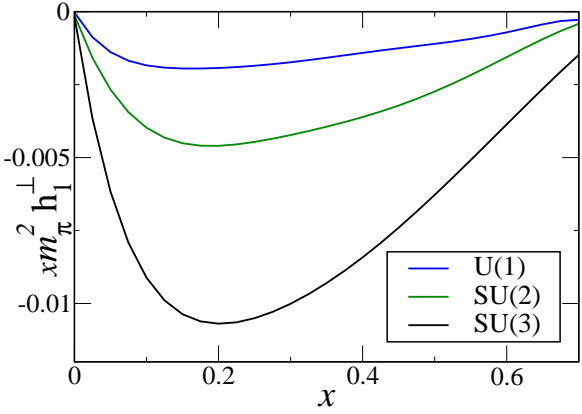}
\end{minipage}
\caption{\scriptsize The pion Boer-Mulders function,  $xm_\pi^2 h_1^{\perp,(1)}(x)$ vs. $x$ calculated by means
of the relation to the chirally-odd GPD $\mathcal{H}_{1}^{\pi}$ for
a SU(3), SU(2), U(1) gauge theory.
 \label{fig:The-eikonal-phase}}
\end{center}
\end{figure}
In this section we  use the eikonal model for the lensing
function together with the spectator model for the GPD $H_{1}^{\pi}$
 to present predictions of the relation (\ref{eq:Relation}) 
for the first moment of the pion Boer-Mulders
function $h_{1}^{\perp(1)}$.

We  start by  fixing the model parameters 
in (\ref{eq:H1Pi}). We encounter six free model parameters $m_{s}$, $m_{q}$, $\Lambda$,
$\lambda$, $g_{\pi}$ and $n$ that we need to determine by fitting
to pion data. In order to do so we determine the chiral-even GPD $F_{1}^{\pi}$
(for definition and notation see Ref.~\cite{Meissner:2008ay}) in
the spectator model,
\begin{eqnarray}
\hspace{-0.5cm} F_{1}^{\pi}(x,0,-\vec{\Delta}_{T}^{2}) \hsp &=&\hsp\frac{g_{\pi}^{2}}{2(2\pi)^{2}}\mathrm{e}^{2\lambda^{2}\Lambda^{2}}\left(\frac{(1-x)\Lambda^{2}}{\vec{D}_{T}^{2}+\tilde{\Lambda}^{2}(x)}\right)^{2n-2}
\label{eq:F1pi}
\\
&&\hspace{-2.5cm}\times\int_{0}^{2\pi}d\varphi\int_{0}^{1}dz\frac{z^{2n-3}\left[1-2z-z\frac{x(1-x)m_{\pi}^{2}-2m_{s}m_{q}}{\vec{D}_{T}^{2}+\tilde{\Lambda}^{2}(x)}\right]\mathrm{e}^{-\frac{2\lambda^{2}(\vec{D}_{T}^{2}+\tilde{\Lambda}^{2}(x))}{(1-x)z}}}{\left[1-4z(1-z)\frac{\vec{D}_{T}^{2}}{\vec{D}_{T}^{2}+\tilde{\Lambda}^{2}(x)}\cos^{2}\varphi\right]^{n}}. \nnn
\end{eqnarray}
When integrated
over $x$, the GPD reduces to the pion form factor $F^{\pi^{+}}(Q^{2})=-F^{\pi^{-}}(Q^{2})$.
An experimental fit of the pion form factor to data is presented in
Refs.~\cite{Blok:2008jy,Huber:2008id}, and up to $Q^{2}=2.45\,\mathrm{GeV^{2}}$
is displayed by the monopole formula
$F_{\mathrm{fit}}(Q^{2})=(1+1.85\, Q^{2})^{-1}$. This procedure is
expected to predict the $t$-dependence of the chirally-odd GPD $H_{1}^{\pi}$
reasonably well up to $Q^{2}=2.45\,\mathrm{GeV^{2}}$. In order to
fix the $x$-dependence of $H_{1}^{\pi}$ we fit the collinear limit
$F_{1}^{\pi}(x,0,0)$ to the valence quark distribution in a pion. 
A parameterization for $F_{1}^{\pi}(x)$
 was given by GRV in Ref.~\cite{Gluck:1991ey} at a scale $\mu^{2}=2\,\mathrm{GeV^{2}}$.  Reasonable agreements of the form factor- and collinear limit of Eq.~(\ref{eq:F1pi})
with the data fits are found for the parameters $m_{q}=0.834\,\mathrm{GeV}$,
$m_{s}=0.632\,\mathrm{GeV}$, $\Lambda=0.067\,\mathrm{GeV}$, $\lambda=0.448\,\mathrm{GeV}$,
$n=0.971$, $g_{\pi}=3.604$. Details of the fitting procedure for this and an
analogous calculation for the Sivers function will be presented in a future
publication~~\cite{gam_schle_sivers}.

With the predicted GPD $H_{1}^{\pi}$ and the lensing function $\mathcal{I}^{i}(x,\vec{b}_{T})\equiv b_{T}^{i}/|b_{T}| \mathcal{I}(x,|b_{T}|)$
as input we use (\ref{eq:Relation}) to give a prediction
for the valence contribution to the first $k_{T}$-moment of the pion
Boer-Mulders function, 
\begin{eqnarray}
m_{\pi}^{2}h_{1}^{\perp(1)}(x)=2\pi\int_{0}^{\infty}db_{T}\, b_{T}^{2}\mathcal{I}(x,b_{T})\frac{\partial}{\partial b_{T}^{2}}\mathcal{H}_{1}^{\pi}(x,b_{T}^{2}).
\end{eqnarray}
Numerical results for $m_{\pi}^{2}h_{1}^{\perp(1)}(x)$ are presented 
in Fig.~\ref{fig:The-eikonal-phase} for a U(1), SU(2) and SU(3)
gauge theory. One observes that all results are negative which reflects
the sign of the lensing function. It was argued in Ref.~\cite{Burkardt:2005hp} 
that a negative sign of the lensing functions indicates attractive
FSIs. We find that this is  valid in an
Abelian perturbative model as well as 
our non-perturbative model in a non-Abelian gauge theory. 
The magnitude of the SU(3) result
is about $0.01$, while the SU(2) result and U(1) result are
smaller.   One observes a growth
of the pion Boer-Mulders function with $N_{c}$. 
Similar growth was also predicted
by a model-independent large $N_{c}$ analysis
for the nucleon~\cite{Pobylitsa:2003ty}, 
though with different leading order behavior.
A phenomenological calculation of  the ratio $k_\perp h_{1}^{\perp\pi}(x,k_\perp)/m_\pi F_1^\pi(x,k_\perp)$ was carried out  in a perturbative,  Abelian,   
one gluon exchange approximation 
and used  to estimate the $\cos2\phi$ AA in  $\pi^- p$ Drell-Yan 
scattering~\cite{Lu:2004hu}.
Similar effects were seen as compared to $p\bar{p}$ Drell Yan 
scattering~\cite{Lu:2005rq,Gamberg:2005ip,Lu:2006ew,Barone:2008tn}. 
It will be useful to 
study  the  dependence of the $\cos2\phi$ AA on color degrees of freedom
from  this non-perturbative approach.  So far the pion Boer-Mulders function is an unknown function
 but  may be accessible  
from a proposed pion-proton Drell-Yan experiment 
by the COMPASS collaboration. 
If  a pion Boer-Mulders function is extracted
from such an experiment our analysis can be used to  quantitatively test 
the GPD - TMD relation (\ref{eq:Relation}). As a comparison, an extraction of another T-odd parton distribution,
the proton Sivers function $f_{1T}^{\perp(1)}$, from SIDIS data measured
at HERMES and COMPASS~\cite{Anselmino:2008sga,Arnold:2008ap} reveals an effect of the magnitude of about
$0.04$. A similar calculation using eikonal methods for the
proton Sivers function will be reported elsewhere~\cite{gam_schle_sivers}.

\section{Conclusions}

In this paper we examined the FSIs of an active quark 
in a pion which are essential to generate a non-vanishing chirally-odd and
(naive) T-odd parton distribution i.e. the Boer-Mulders function.
We considered a pion in a valence quark configuration and worked in a spectator
framework. The FSIs then were modeled by a non-perturbative
2 $\rightarrow$ 2  scattering amplitude 
which we calculated using eikonal methods. This
treatment sums up all soft gluons between re-scattered quark and
antiquark while taking into account color degrees of freedom
and leads to a more complete description of FSIs
as compared to calculations in the 
perturbative Abelian  one gluon exchange approximation.
We find that under  the kinematical conditions 
of soft gluon
exchange for FSIs,  the Boer-Mulders
function can be split into FSIs modeled by eikonal methods
and a spatial distribution of quarks 
in a plane transverse to the direction of motion.
This spatial distribution is described by a chirally-odd pion 
impact parameter GPD 
which we calculate in the spectator model. Together, both effects, 
i.e. FSI and spatial 
distortion  give a prediction for the first moment of the 
Boer-Mulders function
that can be tested in pion-proton Drell-Yan experiments.
\begin{flushleft}
{\bf Acknowledgments}
\end{flushleft}
We thank D. Boer, S. Brodsky, M. Burkardt, H. Fried, 
G. Goldstein, S. Liuti, 
A. Metz, P.J. Mulders, J.-W. Qiu, O. Teryaev, and H. Weigel
for useful discussions. L. G. is grateful for 
for support from G. Miller and  
 the Institute For Nuclear Theory, University of Washington 
where part of this work was undertaken.  
L.G. acknowledges support from  U.S. Department of Energy under 
contract DE-FG02-07ER41460.  Authored by Jefferson Science Associates, LLC under U.S. DOE Contract No. DE-AC05-06OR23177.
The U.S. Government retains a non-exclusive, paid-up, irrevocable, world-wide license to 
publish or reproduce this manuscript for U.S. Government purposes. 

\bibliographystyle{ref} 
\bibliography{Referenzen}

\end{document}